\documentclass[aps,prl,twocolumn,superscriptaddress,showpacs]{revtex4}
\usepackage[dvips]{graphicx}% Include figure files
\usepackage{dcolumn}% Align table columns on decimal point
\usepackage{bm}% bold math
\begin{document}

\title{Real-Space Imaging of Alternate Localization and Extension of Quasi Two-Dimensional Electronic States at Graphite Surfaces in Magnetic Fields}% Force line breaks with \\

\author{Y. Niimi}
\author{H. Kambara}
\author{T. Matsui}
\affiliation{Department of Physics, University of Tokyo, 7-3-1 Hongo Bunkyo-ku, Tokyo 113-0033, Japan}
\author{D. Yoshioka}
\affiliation{Department of Basic Science, University of Tokyo, 3-8-1 Komaba Meguro-ku, Tokyo 153-8902, Japan}
\author{Hiroshi Fukuyama}
 \email{hiroshi@phys.s.u-tokyo.ac.jp}
\affiliation{Department of Physics, University of Tokyo, 7-3-1 Hongo Bunkyo-ku, Tokyo 113-0033, Japan}

\date{6 December 2006}% It is always \today, today,

\begin{abstract}
We measured the local density of states (LDOS) of a quasi two-dimensional (2D) electron system near point defects on a surface of highly oriented pyrolytic graphite (HOPG) with scanning tunneling microscopy and spectroscopy. Differential tunnel conductance images taken at very low temperatures and in high magnetic fields show a clear contrast between localized and extended spatial distributions of the LDOS at the valley and peak energies of the Landau level spectrum, respectively. The localized electronic state has a single circular distribution around the defects with a radius comparable to the magnetic length. The localized LDOS is in good agreement with a spatial distribution of a calculated wave function for a single electron in 2D in a Coulomb potential in magnetic fields.
\end{abstract}

\maketitle
The quantum Hall (QH) effect is one of the most remarkable and fundamental phenomena in condensed matter physics, where the Hall conductance is quantized to integer multiples of $e^{2}/h$. The essence of the integer QH effect is alternating localization and extension of a two-dimensional (2D) electron gas (2DEG), which is realized when the Fermi energy ($\varepsilon_{F}$) is tuned to one of the Landau levels (LLs) and in between the adjacent LLs, respectively~\cite{yoshioka}. In the localized states, it is commonly believed that the 2DEG is trapped around impurities or sample edges running along the equipotential lines with an approximate width of the magnetic length $l_{B}$ ($=\sqrt{\hbar/eB}$ where $B$ is magnetic field). The latter case is called the ``QH edge state.'' However, it is generally difficult to observe such localized states on nanometer scale in real space because the 2D electron systems (2DESs) are usually formed at heterojunctions several hundreds of nanometers below semiconductor surfaces.

Recently, Morgenstern \textit{et al}.~\cite{morgenstern1} studied the localized and extended states in the absorbate-induced 2DES at a cleaved InAs(110) surface with submonolayer iron deposition using the scanning tunneling microscopy (STM) and spectroscopy (STS) techniques. They observed the drift state with a width of $l_{B}$ in the disorder potential, which indicates that the localization in the QH state is described by the conventional single-particle drift motion of electrons. However, probably due to the large number of impurities in their sample, the contrast between the localized and extended distributions of the electronic states is somewhat ambiguous. On the other hand, in high mobility samples, it was claimed that the localization is dominated by Coulomb interactions even in the integer QH regime~\cite{cobden,yacoby}.

The aim of this work is to verify the hypothesis of the localized and extended states of the QH effect and reveal the nature of the localization in the QH state in real space with STM/STS. We studied a quasi 2D electron and hole system at a surface of highly oriented pyrolytic graphite (HOPG) which can be easily accessed with STM/STS. Recent transport~\cite{kopelevich,matsui1} and STS~\cite{matsui2} measurements revealed that the 2D nature of the electronic properties of HOPG is much stronger than that of bulk (single crystal) graphite because of its much higher stacking-fault density. Moreover, the observed Hall resistance plateau~\cite{kopelevich,matsui1} indicates the probable QH state in this material. Since HOPG has very few impurities ($\leq$ 10 ppm), it is a clean 2DES with high mobility ($\mu \sim 10^{5}$ cm$^{2}$/V$\cdot$s)~\cite{geim1}.

\begin{figure}
\begin{center}
\includegraphics[width=6.5cm]{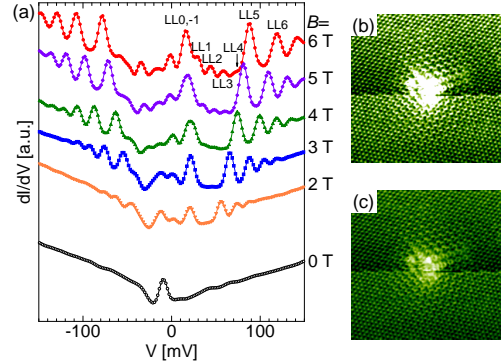}
\caption{(a) Tunnel spectra for a clean HOPG surface at $T=30$ mK in several different magnetic fields perpendicular to the graphite basal plane ($V=180$ mV, $I=0.2$ nA). Each spectrum is vertically shifted for clarity. (b), (c) STM images with the same surface point defects at the center represented with different higher (b) and lower (c) contrasts ($8 \times 8$ nm$^{2}$, $V=180$ mV, $I=0.2$ nA, $T=30$ mK).
}
\label{fig1}
\end{center}
\end{figure}

\begin{figure*}
\begin{center}
\includegraphics[width=11.5cm]{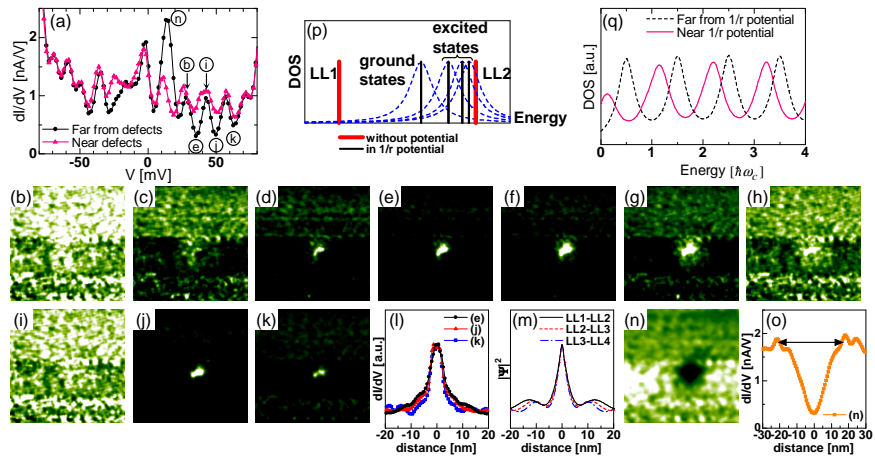}
\caption{(a) Tunnel spectra averaged over $20 \times 20$ nm$^{2}$ centered on the defects (triangle) and far away ($>30$ nm) from them (circle) at a fixed field of 6 T ($V=180$ mV, $I=0.2$ nA, $T=30$ mK). (b)-(k), (n) d$I/$d$V$ images over $80 \times 80$ nm$^{2}$ around the defects at various bias voltages [(b) $28$, (c) $32$, (d) $33$, (e) $35$, (f) $37$, (g) $39$, (h) $41$, (i) $43$, (j) $50$, (k) $63$, and (n) $14$ mV]. (l) Typical cross sections of the d$I/$d$V$ images at the valley energies. The vertical axis is normalized by the central peak height. (m) The energy dependence of calculated wave functions in the $1/r$ potential at 6 T. (o) A typical cross section of the d$I/$d$V$ image at the peak energy of LL$0,-1$. The arrow indicates the approximate diameter (35 nm) of the confining potential due to the defects. (p) The vertical lines represent a schematic energy-level scheme of 2D single electron states in the $1/r$ potential and LLs without the potential. The broken curves are DOS associated with those levels with finite widths determined by electron lifetimes. (q) Calculated DOS for 2DEG at 6 T averaged over areas of $30 \times 30$ nm$^{2}$ centered on the $1/r$ potential (solid line) and far away (broken line).}
\label{fig2}
\end{center}
\end{figure*}

In this Letter, we present differential tunnel conductance (d$I/$d$V$) images obtained around point defects at the HOPG surface. The preliminary data have been shown in Ref. 7, but more detailed measurements and new analyses are given here. The d$I/$d$V$ images, i.e., the local density of states (LDOS) mappings, revealed clear distinction between localized and extended distributions of electronic wave functions depending on bias voltage. At an energy in between the adjacent LLs, a circular distribution of the LDOS with a radius comparable to $l_{B}$ was observed near the point defects. By comparing with simple theoretical calculations, we show that the functional form of disorder potential is crucial for details of the localized state.

The present STM/STS measurements were performed at temperatures below 30 mK and in magnetic fields up to 6 T using an ultra low temperature STM~\cite{ult-stm}. The HOPG sample~\cite{HOPG} is synthesized by chemical vapor deposition and subsequent heat treatment under high pressures. The sample was cleaved in air and then quickly loaded into an ultra high vacuum chamber of the STM. The d$I/$d$V$ curves and images were taken by the lock-in technique with a bias modulation $V_{\rm mod}$ of 1.0 mV at a frequency of 412 Hz.

Figure 1(a) shows typical tunnel spectra measured at the HOPG surface without nearby defects in several different magnetic fields. Clear LL peaks are observed as in a previous work~\cite{matsui2}, where intervals of the peaks expand in proportion to $B$ and the peak heights increase with increasing field. These peaks at positive and negative bias voltages are associated with the electron and hole LLs, respectively. The electron LL peaks are labeled such as LL1, LL2, $\cdots$ from the lower levels to higher ones. The peak structures here are a factor of 4 more pronounced compared to those in Fig. 3 of Ref. 7 presumably because of  the smaller effective thickness~\cite{note1}. On the other hand, a relatively pronounced peak just above $\varepsilon_{F}$ at finite fields has an almost field independent energy ($=+20$ meV). It originates from the $n=0$ (electron) and $n=-1$ (hole) LLs which are characteristic of the graphite lattice structure~\cite{Dresselhaus,nakao}. This field independent peak is denoted as LL$0,-1$. Note that, in zero magnetic field, there is a small peak at an energy ($=-10$ meV) just below $\varepsilon_{F}$. This is likely due to a local electrostatic potential induced by the STM tip, as was reported in previous STM works on semiconductor~\cite{morgenstern1,morgenstern2} and graphite surfaces~\cite{matsui2}.

In Figs. 1(b) and (c), we show two STM images, which are represented with different contrasts, around point defects we found on the HOPG surface. One can see a clear three fold scattering pattern from the defects and a $\sqrt{3} \times \sqrt{3}$ superstructure with a decay length of about 3 nm. These patterns are consistent with calculations for a few adjacent atomic defects on a single graphite sheet~\cite{mizes} and with the experimental observation~\cite{superstructure}. The same superstructure has also been observed near graphite step edges~\cite{niimi1}. Although one may expect to extract an atomic defect structure from Fig. 1(c), it is difficult to determine it unambiguously only from the STM image. 

\begin{figure*}
\begin{center}
\includegraphics[width=11cm]{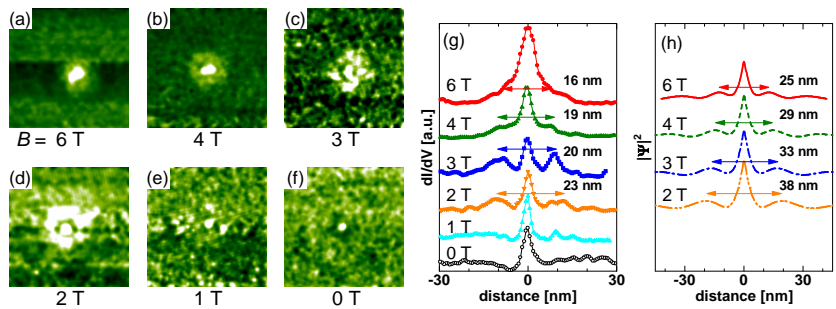}
\caption{(a)-(f) The magnetic field dependence of d$I/$d$V$ images ($80 \times 80$ nm$^{2}$, $I=0.2$ nA, $T=30$ mK) near the point defects at the valley energies in between LL1 and LL2; (a) $B=6$ T ($V=35$ mV), (b) 4 T (32 mV), (c) 3 T (30 mV), (d) 2 T (28 mV), (e) 1 T (26 mV), (f) 0 T (25 mV). (g) Typical cross sections of the d$I/$d$V$ images in the different fields. The arrows show the approximate diameters of the LDOS rings. (h) The field dependence of calculated wave functions at the valley energies in between LL1 and LL2 in the $1/r$ potential.}
\label{fig3}
\end{center}
\end{figure*}

Next, we show STS data near the point defects at $B=6$ T. As is shown in Fig. 2(a), the LL peak heights except LL$0,-1$ are almost the same between the spectrum averaged over $20 \times 20$ nm$^{2}$ centered on the defects and the one far away ($>30$ nm) from them. Meanwhile, at the valley energies between the adjacent two LLs above $\varepsilon_{F}$, the LDOS near the defects is obviously higher than that far from the defects. The d$I/$d$V$ mappings show such a difference in a more dramatic manner (Figs. 2(b)-(k)). Note that the same contrast is used for all the images and that the scan area is a hundred times larger than Figs. 1(b) and (c). At the peak energy of LL1 (Fig. 2(b)), the electronic state is relatively extended. With increasing energy, it gradually localizes at the defects (Figs. 2(c) and (d)). At the valley energy in between LL1 and LL2, a circular LDOS distribution with a radius ($\sim 5$ nm) of a half of $l_{B}$ is clearly seen around the defects (Fig. 2(e)). The circular distribution gradually disintegrates with increasing energy up to the LL2 (Figs. 2(f)-(i)). The same evolution is observed at energies in between LL2 and LL3 (LL3 and LL4). These results are consistent with the alternating localization and extension of the 2DES in the QH state. And this fact suggests the possible QH state at the HOPG surface.

Interestingly, for negative bias voltages below $-35$ mV, such clear localized states at the valley energies were not observed as in Fig. 2(a) where the spectra either near or far from the defects are identical. Let us argue this energy sign dependence in the following. The LLs at positive energies originate only from the electron band. On the other hand, the LLs in an energy range between $-80$ and $-15$ meV in the case of 6 T (see Fig. 2 of Ref. 7) consist both of the electron and hole bands. They are located near the $K$ and $H$ points in the hexagonal Brillouin zone of graphite, respectively. Such coexistence of the two different kinds of carriers with different effective masses and cyclotron radii could play a destructive role in the formation of localized states. This hypothesis is consistent with the recent transport measurements~\cite{kopelevich,matsui1}. The measured Hall resistance in HOPG shows a QH-effect-like plateau in the field range between 4 and 6 T where only the second lowest electron LL besides LL$0,-1$ lies across $\varepsilon_{F}$. But it does not show marked plateau structures below 4 T where both the electron and hole LLs lie across $\varepsilon_{F}$. Therefore, our STM/STS data indicate that at least the electron LLs above $\varepsilon_{F}$ contribute to the formation of the localized state.

It is interesting to note that the LDOS at the peak energy of LL$0,-1$ is considerably reduced near the defects. In the previous work~\cite{matsui2}, it was reported that this peak is very sensitive to the surface potential. Thus, the d$I/$d$V$ image taken at this energy should map the defect potential distribution (Fig. 2(n)). Actually, the cross section of the image has a hollow shape with an approximate diameter of 35 nm (Fig. 2(o)) and is nearly independent of magnetic field.

Figures 3(a)-(f) show the spatial distributions of the localized states in between LL1 and LL2 at several different magnetic fields~\cite{note2}. At 6 T, the electronic state is localized not only just on the defects but also around them as a ring with a radius ($=8$ nm) slightly less than $l_{B}$. Note that Fig. 3(a) is the same image as Fig. 2(e) but is shown in a slightly different contrast. With decreasing field until 2 T, the amplitude of the LDOS just on the defects decreases, while the LDOS ring around the defects expands (see Fig. 3(g)). Such a field dependence is also observed near several other defects at the HOPG surface (Fig. 4)~\cite{niimi2}. This behavior cannot be explained, for example, by the weak-localization which should be suppressed with increasing field due to the time-reversal symmetry breaking. Thus, the localized states observed here are attributable only to the drift states of quasi-2DEG in magnetic fields. Below 1 T, we did not observe such a clear LDOS ring (Fig. 3(e)). This is presumably because the spatial extent of the localized state at that field exceeds the phase coherent length in the system.

\begin{figure}
\begin{center}
\includegraphics[width=8cm]{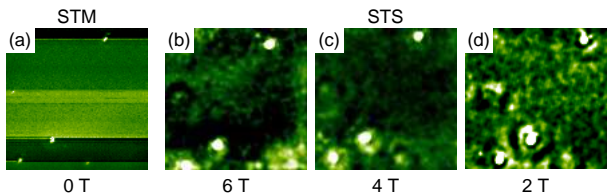}
\caption{(a) STM image near several defects on the HOPG surface at zero field ($100 \times 100$ nm$^{2}$, $V=180$ mV, $I=0.2$ nA, $T=30$ mK). (b)-(d) The magnetic field dependence of d$I/$d$V$ images ($100 \times 100$ nm$^{2}$, $I=0.2$ nA, $T=30$ mK) near the same defects as shown in (a) at the valley energies in between LL1 and LL2; (b) $B=6$ T, (c) 4 T, (d) 2 T.}
\label{fig4}
\end{center}
\end{figure}

We calculated eigenenergies and eigenfunctions of 2DEG in a Coulomb-type potential $U(r)= -\alpha/r$ in magnetic fields by solving the Schr\"{o}dinger equation in order to compare the measured localized states with them. The degeneracy of the LLs is lifted by the electrostatic potential (Fig. 2(p)). For the case of $1/r$ potential, we found that the ground state at each LL has zero angular momentum. The reason for this is to minimize the potential energy for electron by maximizing a wave function amplitude at the origin. It is in clear contrast to the case of harmonic potential~\cite{fock}, in which the maximum of the lowest energy wave function at any LL except LL0 is not at origin but forms a ring around it. The observed LDOS can be explained fairly well only by the wave functions of the ground states in the 1/r potential. However, a quantitatively better agreement with the experiment in terms, for example, of the radial distributions of the satellite peaks was obtained by taking account of smaller contributions from the excited states with finite angular momenta. Note that the electron charging effects should not be observable in our STS measurements at least at positive bias voltages since the probability of electron double occupancy in the same localized state is negligibly small. This was judged from the fact that the electron lifetimes ($\approx$ 10$^{-12}$ s) in the localized states estimated from the LL peak widths ($\approx$ 3 meV) in the tunnel spectra are 3 orders of magnitude shorter than the average time interval ($\approx$ 10$^{-9}$ s) between the two successive events of electron tunneling from the STM tip. We assumed the same peak width for the exited states in the calculations (Fig. 2(p)).

In Fig. 2(m), we plot calculated wave-functions at the valley energies in between LL1 and LL2, LL2 and LL3, and LL3 and LL4 at 6 T. Note that $\alpha$, the coefficient of the $1/r$ potential, was determined as 90 meV$\cdot$nm from the apparent energy shifts of the LLs near the defects (see Figs. 2(a) and (q))~\cite{niimi2}. The width of the central peak ($\sim$ 0.5 $l_{B}$) slightly shrinks with increasing energy, which is consistent with the experimental results (Fig. 2(l)). In Fig. 3 (h), we show the calculated wave functions at the valley energies in between LL1 and LL2 in several fields. They have not only the central peaks but also satellite ones. The diameter of the satellite peak increases with decreasing field. These qualitatively reproduce the measured LDOS distributions in the d$I/$d$V$ images (Figs. 3(a)-(d) and (g)). However, there are a few quantitative discrepancies between the present experiments and calculations. For example, the calculated diameter of the satellite peak is larger than that of the measured one by a factor of 1.5. This could be improved by taking account of the quasi two-dimensionality in the calculations.

In summary, we succeeded in the first clear visualization of the alternating electron localization and extension depending on energy near the point defects at the surface of highly oriented pyrolytic graphite (HOPG) in real space by the STM/STS measurements. The single circular distribuitons in the localized states are semi-quantitatively consistent with the calculated wave functions for 2DEG in the $1/r$ potential in magnetic fields. These results reveal that the functional form of potential plays an important role in the formation of the localized states. This work also indicates the possibility of the quantum Hall state at the surface of HOPG which has been suggested by the recent transport measurements.

We thank M. Tsukada, K. Tagami, H. Aoki and M. Ogata for valuable discussions. The authors are grateful to C. Winkelmann for useful comments on this manuscript. This work was financially supported by Grant-in-Aid for Scientific Research on Priority Areas (Grant No. 17071002) from MEXT and ERATO Project of JST. Y.N. and T.M. acknowledge the JSPS Research program for Young Scientists.

%\newpage %Just because of unusual number of tables stacked at end
%\bibliography{revised_manuscript(Niimi_et_al)}%

\begin{thebibliography}{00}
\bibitem{note1}
The Green's function calculations show that the surface LDOS of thinner graphite has more pronounced and complicated peak structures (see Ref. 7).
\bibitem{note2}
Preliminary results of the localized states have been reported previously in Ref. 7 without any quantitative analysis.
\bibitem{niimi2}
Further details will be published elsewhere.
\end{thebibliography}

\end{document}